\documentclass[twocolumn, aps, prb, floatfix]{revtex4}
\usepackage{graphicx}

\begin{document}
\title{How two-dimensional brick layer J-aggregates differ from linear ones: excitonic properties and line broadening mechanisms}
\author{Arend~G.~Dijkstra$^{1,2}$}
\author{Hong-Guang Duan$^{2}$}
\author{Jasper Knoester$^3$}
\author{Keith~A.~Nelson$^1$}
\author{Jianshu Cao$^1$}
\affiliation{1) Department of Chemistry, Massachusetts Institute of Technology, 77 Massachusetts Avenue, Cambridge, MA 02139, United States of America}
\affiliation{2) Max Planck Institute for the Structure and Dynamics of Matter, Luruper Chaussee 149 Bldg. 99, 22761 Hamburg, Germany}
\affiliation{3) Zernike Institute for Advanced Materials, Nijenborgh 4, 9747 AG Groningen, The Netherlands}

\begin{abstract}
We study the excitonic coupling and homogeneous spectral line width of brick layer J-aggregate films. We begin by analysing the structural information revealed by the two-exciton states probed in two-dimensional spectra. Our first main result is that the relation between the excitonic couplings and the spectral shift in a two-dimensional structure is different (larger shift for the same nearest neighbour coupling) from that in a one-dimensional structure, which leads to an estimation of  dipolar coupling in two-dimensional lattices. We next investigate the mechanisms of homogeneous broadening - population relaxation and pure dephasing - and evaluate their relative importance in linear and two-dimensional aggregates. Our second main result is that pure dephasing dominates the line width in two-dimensional systems up to a crossover temperature, which explains the linear temperature dependence of the homogeneous line width. This is directly related to the decreased density of states at the band edge when compared with linear aggregates, thus reducing the contribution of population relaxation to dephasing. Pump-probe experiments are suggested to directly measure the lifetime of the bright state and can therefore support the proposed model.
\end{abstract}
\date\today
\maketitle

\section{Introduction}
Organic molecules are promising candidates for the next generation of electronic devices and for solar energy conversion.\cite{Gelinas.2013.science.343.512, Bakulin.2012.science.335.1340, Haedler.2014.chemeur.20.11708, Jailaubekov.2013.natmat.12.66, Ley.2014.jacs.136.7809, Roberts.2012.jacs.134.6388}
Among these, assemblies of dye molecules in the form of J-aggregates have attracted attention for their special optical properties.\cite{Jelley.1936.nature.138.1009,Scheibe.1936.angew.49.563} These are understood from delocalization of the exciton formed upon the absorption of light over tens to hundreds of monomers.\cite{Knapp.1984.chemphys.85.73} Because the dominant resonant transfer interactions between molecules are negative in a J-aggregate, the optically bright state is found at the bottom of the band, leading to a redshift of the absorption peak compared to a single chromophore. Other properties resulting from this exciton delocalization are superradiance and a hidden level structure at the band edge.\cite{Bednarz.2003.prl.91.217401, Augulis.2010.jpcl.1.2911} The exciton delocalization in linear aggregates is reflected in the pump-probe spectrum\cite{Juzeliunas.1988.zpd.8.379, Bakalis.1999.jpcb.103.6620} and in the two-dimensional optical spectrum.\cite{Han.2013.jcp.139.034313}

Most early studies focused on J-aggregates for which the optical properties can be explained with a model of a linear aggregate, which self-assemble in solution and are often studied at low temperature in a glass environment.\cite{Fidder.1991.jcp.95.7880} Over the past years there has been an intense interest in tubular J-aggregates.\cite{Didraga.2002.jpcb.106.11474, Eisele.2012.natchem.4.655, Milota.2013.jpca.117.6007, Eisele.2014.pnas.111.E3367, Stradomska.2010.jcp.133.094701, Chuang.2015.arxiv.1511.01198, Lim.2015.ncomm.6.7755} It is also possible to manufacture two-dimensional thin film J-aggregates of chromophore molecules, which were found to exhibit a redshift in the absorption.\cite{Arias.2012.jpcb.117.4553, Muller.2013.jcp.139.044302} Nonlinear optical experiments produced a two-dimensional spectrum similar to the spectrum of a linear aggregate, consisting of a single pair of positive and negative peaks.\cite{Arias.2012.jpcb.117.4553}

In order to analyse these findings, a model of a truly two-dimensional aggregate must be used,\cite{Moebius.1988.japplphys.64.5138, Valleau.2012.jcp.137.034109} which goes beyond weakly coupled linear aggregates.\cite{Chuang.2014.jpcb.118.7827} In general, the transfer interactions between molecules depend strongly on their relative orientation. This means that the absorption spectrum is sensitive to the details of the molecular arrangement. Therefore, modeling of the spectrum can help in determining the structure. This is particularly helpful in cases where the structure is not known from other measurements.\cite{Arias.2012.jpcb.117.4553}

It is clear from experiment, as well as from consideration of the molecular structure, that the excitons in J-aggregates must couple to their environment. This coupling leads to scattering between exciton states and to pure dephasing. The total dephasing process can be studied by measuring the homogeneous line width in experiments such as photon echos, hole burning or two-dimensional spectroscopy. In particular, the dependence of the homogeneous line width on temperature can be analysed in order to understand the exciton phonon coupling mechanism. Understanding the interaction with the environment is also important to assess how many molecules in the aggregate are entangled upon optical excitation. This entanglement, or delocalization, is the key quantity that causes the interesting properties of these systems. However, interactions with phonons limit the localization size, and determining this quantity as a function of temperature is an important goal. It has been shown that the energy dependent localization size can be extracted from two-dimensional spectra.\cite{Dijkstra.2008.jcp.128.164511}

Studies of the homogeneous line width have been performed on J-aggregates for which linear chain models explain the optical properties.
Fidder et al.\cite{Fidder.1990.cpl.171.529} measured the homogeneous width in PIC-Br for temperatures between 1.5 and 190 K. The temperature dependence is clearly nonlinear and was modeled by coupling of the excitons to three harmonic modes with frequencies of 9 cm$^{-1}$, 305 cm$^{-1}$ and 973 cm$^{-1}$.
Hirschmann and Friedrich\cite{Hirschmann.1989.jcp.91.7988} measured the homogeneous line width in PIC-I for temperatures from 0.35 K to 80 K. The temperature dependence can be fitted by a sum of two exponentials or be explained by a theory that predicts a power law dependence.\cite{Heijs.2005.jcp.123.144507} The homogeneous line shape is well approximated by a Lorentzian for all temperatures.

In contrast to this work, experiments on thin films using two-dimensional optical spectroscopy have found a \underline{linear} scaling of the homogeneous line width with temperature.\cite{Arias.2012.jpcb.117.4553} This suggests that a different mechanism is responsible for the line width. 

In this work, we use an excitonic model of two-dimensional brick layer J-aggregates and study the homogeneous line width as a function of temperature. Our theoretical model is presented in Section II and the results are presented in Section III.  Specifically, we predict the excitonic coupling in molecular aggregates and thus correlate two-dimensional spectra with molecular arrangements. These calculations are the topic of Sections IIIA, B, and C.  Then we consider population relaxation and the exciton lifetime n Sec. IIID and calculate the homogeneous line width as a function of temperature in Section IIIE.  Finally, we analyse experimental measurements in Sec. IV and conclude in Section V. In the appendix, we consider alternative aggregate geometries.

\section{Model} \label{sec:model}
The usual Frenkel exciton model of J-aggregates starts from a single bright optical transition on each molecule. The Hamitonian includes a term which describes local excitation of a molecule with an excitation energy $\epsilon_n$ and a term for the coherent exciton motion from one molecule to the other, and is given in terms of the Pauli creation and annihilation operators $c^\dagger$ and $c$ by
\begin{equation}
  H_S = \sum_n \epsilon_n c_n^\dagger c_n + \sum_{nm} J_{nm} c_n^\dagger c_m.
\end{equation}
In this Hamiltonian, the sums run over all molecules in the aggregate. If the molecules are far enough apart, the electrostatic interaction between them can be approximated by dipole dipole coupling, which gives
\begin{equation}
  J_{nm} = C \frac{\mu_n \cdot \mu_m - 3 (\mu_n \cdot \hat r_{nm})(\mu_m \cdot \hat r_{nm})}{r_{nm}^3}.
\end{equation}
Here, $\mu_n$ is the transition dipole vector of molecule $n$, $\vec r_{nm} = \vec r_n - \vec r_m$ is the relative position vector, $r_{nm} = |\vec r_{nm}|$ is the distance and $\hat r_{nm} = \vec r_{nm} / r_{nm}$. $C$ is a constant that scales the magnitude of the coupling and includes possible rescaling effects due to vibrations.\cite{Mueller.2013.jcp.138.064703} Here, we will use transition dipole coupling for all pairs of molecules. For nearest neigbours, a better understanding of the coupling can be obtained from quantum chemical calculations.\cite{Lee.2013.jpca.117.11072, Kistler.2013.jpcb.117.2032}

Each molecule in the aggregate is influenced by a different local environment. This leads to static disorder in the site energies $\epsilon_n$, which are different for each aggregate in the ensemble. This, in turn, leads to localization and a distribution of effective sizes of the exciton. Furthermore, the excitons in the aggregate interact with phonons in the surrounding material. Their dynamic effect is usually modeled as a reservoir of harmonic oscillators, which are described by the bath Hamiltonian $H_B$. The interaction of these oscillators with the electronic excitations is assumed to be
\begin{equation}
  H_\mathrm{SB} = \sum_n X_n c_n^\dagger c_n,
\end{equation}
where the effective bath coordinate is to be thought of as the sum of couplings to individual bath modes, which can be written as $X_n = -\sum_\alpha g_{n\alpha} x_\alpha$. Here, $x_\alpha$ denote the coordinates of the bath modes, while $g_{n\alpha}$ are their coupling constants to the system. The linear dependence on the bath coordinate can be thought of as a lowest order expansion in the coordinate. The properties of the system bath interactions are determined by the correlation functions $\langle X_n(t) X_m(0)\rangle$. We will make the usual but not completely general assumption that the fluctuations on each site are uncorrelated and that their correlation function is the same on each site, $\langle X_n(t) X_m(0)\rangle = \delta_{nm} L(t)$.

The system Hamiltonian (for each realization of the static disorder) can be diagonalized to give the exciton states $\phi_{k}$, which we choose to be real, and energies $E_k$, such that $H_S = \sum_k E_k c_k^\dagger c_k$. The wave functions relate the exciton basis to the site basis by the equation $c^\dagger_k = \sum_n \phi_{kn} c^\dagger_n$. In the same exciton basis, the system bath interaction can be written as $H_\mathrm{SB} = H_\mathrm{SB}^{(0)} + H_\mathrm{SB}'$, with the diagonal fluctuations
\begin{equation}
H_\mathrm{SB}^{(0)} = \sum_{kn} \phi_{kn}^2 X_n c_k^\dagger c_k,
\end{equation}
and the off-diagonal fluctuations that couple two different eigenstates
\begin{equation}
  H_\mathrm{SB}' = \sum_{q\neq k, n} \phi_{kn} \phi_{qn} X_n c_k^\dagger c_q.
\end{equation}

In modified Redfield theory\cite{Cho.2005.jpcb.109.10542, Abramavicius.2010.jcp.133.184501}, the diagonal fluctuations are treated exactly, while the off-diagonal fluctuations are included in second order perturbation theory.
The zero order Hamiltonian $H_S + H_\mathrm{SB}^{(0)} + H_B$ does not couple the exciton states. Therefore, the absorption spectrum for this Hamiltonian is simply the sum of contributions from each exciton state. In this case, where the system Hamiltonian commutes with the system bath interaction, the linear and nonlinear response functions can be calculated analytically with the cumulant expansion. For the linear absorption in the time domain, we find
\begin{equation} \label{lin0}
  A^{(0)}(t) = \sum_k |\mu_k|^2 e^{-i E_k t - g_k(t)},
\end{equation}
where $\mu_k$ is the transition dipole from the ground state to exciton state $k$. The spectrum $A^{(0)}(\omega)$ is given as the Fourier transform of $A^{(0)}(t)$. The line shape function for each exciton state is given by $g_k(t) = g(t) / N_k$, where $N_k = 1/\sum_n \phi_{kn}^4$ is the inverse participation ratio\cite{Cho.2005.jpcb.109.10542}. The line shape function for a single site is defined as
\begin{equation}
g(t) = \int_0^t dt_1 \int_0^{t_1} dt_2 L(t_2).
\end{equation}

If the harmonic bath is interpreted in the continuum limit, the correlation function can be expressed in terms of the spectral density $J(\omega)$ (although, in principle, the spectral density can also contain delta functions which describe discrete modes). The quantum correlation function $L(t)$ is given in terms of the spectral density by the expression
\begin{equation} \label{cfj}
L(t) = \frac{1}{\pi} \int_0^\infty d\omega J(\omega) (\coth \frac{\beta\omega}{2} \cos \omega t - i \sin \omega t).
\end{equation}

The homogeneous line width, which can be measured in photon echos or two-dimensional spectra can now be explained by two broadening mechanisms. First, there is the pure dephasing contribution contained in Eq.~\ref{lin0}. Second, there will be a contribution from $H_\mathrm{SB}'$. In second order perturbation theory with the Markov and secular approximations, this term will lead to dephasing given as the sum of population relaxation rates. This contribution can be termed dephasing from population relaxation. 
The perturbative treatment of $H_\mathrm{SB}'$, which is known from experiment to be weak in certain linear J-aggregates,\cite{Heijs.2005.jcp.123.144507} leads to scattering between exciton eigenstates. The scattering rate between eigenstate $q$ and $k$ is given by
\begin{equation}
  W_{kq} = \sum_n \phi_{kn}^2 \phi_{qn}^2 J(|\omega_{kq}|) n(\omega_{kq}),
\end{equation}
where the sum runs over all molecules in the aggregate, $\omega_{kq} = E_k - E_q$, $J$ is the spectral density and $n(\omega_{kq}) = \bar n(\omega_{kq})$ for $\omega_{kq} > 0$ and $n(\omega_{kq}) = \bar n(-\omega_{kq})+1$ for $\omega_{kq} < 0$, with $\bar n(\omega) = (\exp(\omega/k T) - 1)^{-1}$ the Bose Einstein distribution. The resulting dephasing rate of exciton state $k$ is given by $\Gamma_k = (1/2) \sum_{q \neq k} W_{qk}$. Then, the homogenous absorption line with both pure dephasing and dephasing from population relaxation is given by
\begin{equation} \label{lin}
    A(t) = \sum_k |\mu_k|^2 e^{-i E_k t - g_k(t) - \Gamma_k t}
\end{equation}

Note that we neglect the radiative life time in this expression, which normally gives a negligible contribution to the linewidth. Also, in this work we do not include an ensemble of localization sizes causes by the presence of static disorder. Finally, it should be observed that the homogeneous line width is not equal to the sum of dephasing rates if multiple transitions overlap in the spectrum.

\section{Results}

\subsection{Molecular arrangement and resonant transfer interactions} \label{sec:struct}

\begin{figure*}[t]
 \includegraphics{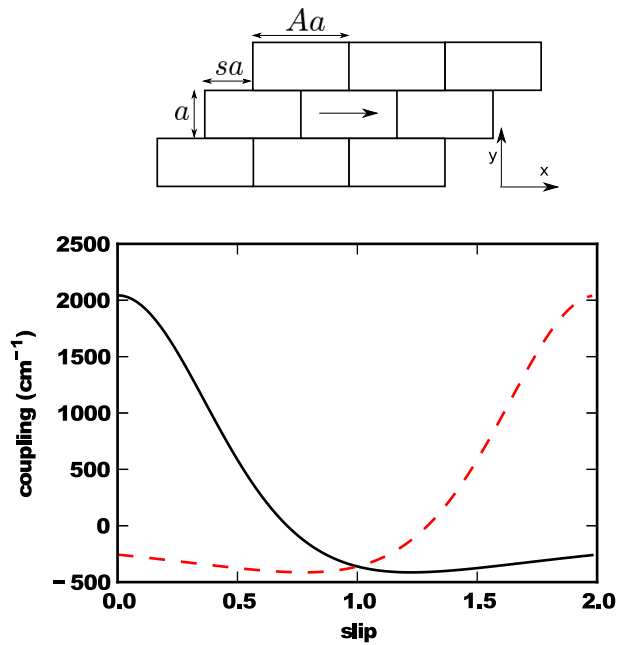}
\caption{\label{fig:coupling} Top: Cartoon of the brick layer model with aspect ratio $A$, slip $s$ and molecular size $a$. (transition dipole indicated by the arrow). Bottom: Coupling between molecules above each other to the right (solid) and left (dashed) as a function of the slip. Note that for a slip of 1.0 (half the unit cell) both couplings are negative, leading to J-coupling for all three nearest neighbours. The aspect ratio is $A=2$ and parameters are chosen such that the coupling in the horizontal direction is kept constant at $J=-511 \mathrm{cm}^{-1}$.}
\end{figure*}

We assume that the molecules are placed on a brick layer lattice\cite{Moebius.1988.japplphys.64.5138, Valleau.2012.jcp.137.034109} with aspect ratio $A$ and slip $s$, see figure \ref{fig:coupling}. The grid has $N_x$ molecules in the x-direction and $N_y$ in the y-direction. These sizes should not be interpreted as the physical size of the aggregate, but as the number of molecules over which an exciton is delocalized. Different geometries can then be obtained by varying the slip, while we assume a constant aspect ratio of $A=2$, which is a reasonable number for molecules typically used to form thin films. Note that with this choice, a slip of 1.0 (half a unit cell) corresponds to a square lattice with dipoles oriented at 45 degrees with respect to the lattice vectors. We note that, for the small aggregates considered in this paper, the choice of boundary conditions is important. We limit our study to the boundary conditions shown in Fig.~\ref{fig:coupling}. In the appendix, we consider the value of $A=3$. Other arrangements, for example herringbone structures with two molecules per unit cell, are outside the scope of this paper. For a constant prefactor $C$ the magnitude of the couplings will change with $s$. This is shown in figure \ref{fig:coupling}. We define the nearest neighbour coupling in the x-direction as $J$. Note that we include all long range couplings in our model as well, and that we don't use periodic boundary conditions. 

In figure \ref{fig:coupling} we observe that the values of $s$ for which an aggregate with negative couplings in both directions is formed are quite limited. In most cases, the coupling in the vertical direction is positive. Because of this combination of negative and positive interactions, the system is not a perfect J-aggregate, in the sense that the bright state is not necessarily at the bottom of the band. We will see that this has observable consequences for the two-dimensional optical spectrum. Negative couplings in both the x and y directions are found around $s=1.0$, which is the structure close to the one assumed in experimental work on brick layer PTCDA aggregates.\cite{Muller.2013.jcp.139.044302}

\begin{figure*}[t]
 \includegraphics{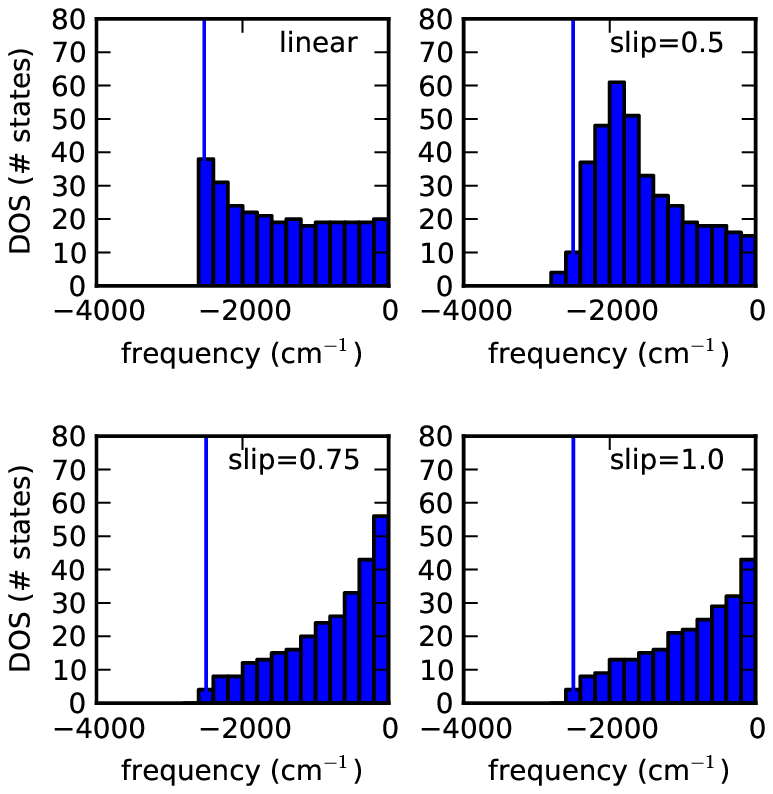}
\caption{\label{fig:dos} Density of states for a linear ($N=625$) and various two-dimensional ($N_x=N_y=25$) aggregates. The bright state is at -2500 cm$^{-1}$ for all cases, as indicated by a vertical line. Couplings were scaled to place the bright state at that position for this number of molecules. The width of the bins is 200 cm$^{-1}$.}
\end{figure*}

In figure \ref{fig:dos} we plot the density of states for a linear aggregate and several two-dimensional aggregates.
For each geometry, we scale the parameter $J$ (or, equivalently $C$), to obtain a shift of approximately 2500 cm$^{-1}$ of the aggregate absorption peak with respect to the absorption peak of the monomer. 
This choice is made to stay close to the interpretation of experimental results, in which the spectral shift upon aggregation can be measured, but the structure of the aggregate (i.e. a two-dimensional bricklayer lattice or a collection of semi-one-dimensional chains) is not always a priori known. In particular, we are interested in comparing with experimental data on BIC aggregates studied in Ref.~\onlinecite{Arias.2012.jpcb.117.4553}, where this shift was observed to be around 2500 cm$^{-1}$. In order to obtain this shift, we set $J=-1050$ cm$^{-1}$ for the linear aggregate, $J=-903$ cm$^{-1}$ for $s=0.5$, $J=-472$ cm$^{-1}$ for $s=0.75$ and $J=-416$ cm$^{-1}$ for $s=1.0$.  We note that, because of differences in the electrostatic environment, the shift does not necessarily directly reflect the excitonic coupling. 

In the linear case, we observe a strong increase of the density at the band edge, reflecting the $1/\sqrt{E}$ scaling in the case of an infinite chain.\cite{Bloemsma.2015.prl.114.156804} The two-dimensional aggregate with $s=0.5$ mirrors this behaviour, but has additional states at lower energy. For larger values of $s$, however, the density of states decreases with decreasing energy, in line with the expected constant behaviour in the limit of an infinite sheet.\cite{Bloemsma.2015.prl.114.156804} We note that the difference in density of states between a linear and two-dimensional aggregate is quite dramatic. For example, for the energy gap between the two lowest-lying states, we find 9.2 cm$^{-1}$ in a 25x25 lattice ($s=1.0$, $J=-416 \, \mathrm{cm}^{-1}$), while it is only 0.45 cm$^{-1}$ in a linear chain with 625 molecules ($J=-1050 \, \mathrm{cm}^{-1}$). (Note that we include long range interactions in the simulations leading to this number.)

\subsection{Two-dimensional spectra}

We now turn our attention to the linear and two-dimensional optical spectra for this model system. For all models the simulation predicts a dominant bright peak in the linear absorption spectra, which, by construction lies about 2500 cm$^{-1}$ below the monomer absorption peak. However, we find that $s=0.5$ and $s=1.0$ are clearly distinguishable when the two-dimensional correlation spectrum is considered. 

Two-dimensional optical spectroscopy is a third-order nonlinear optical technique which correlates the evolution of the electronic state of the system during two time periods, called $t_1$ and $t_3$.\cite{2DESspecialissue} The signal is plotted as a function of the Fourier transforms of these two time periods, with frequencies labeled $\omega_1$ and $\omega_3$. The technique can be used to separate homogeneous broadening, which shows up as the anti-diagonal width of peaks in the two-dimensional plot, from inhomogeneous broadening, which contributes to the diagonal width. It therefore provides a tool to measure the homogeneous line width. When one looks at the real value of a two-dimensional spectrum, both negative peaks, colored in blue, and positive peaks, colored in red, are present. Blue peaks arise from interactions where one excitation is created, while red peaks correspond to processes where, during $t_3$, coherences between one- and two-quantum states (in which two excitation quanta are present in the system) are present. Positive peaks are blue shifted with respect to negative peaks as a consequence of the Pauli exclusion principle. The vertical distance between positive and negative peaks can be used as a ruler from which the exciton localization size can be determined. 

\begin{figure}[t]
 \includegraphics{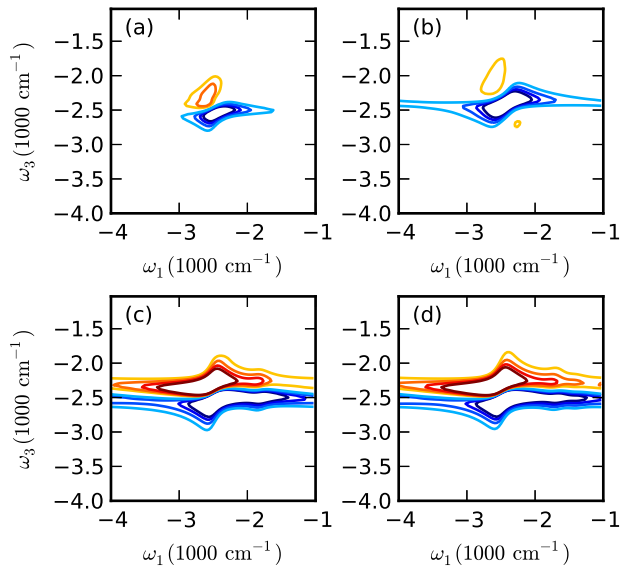}
\caption{\label{fig:spec2D0} Simulated real part of the two-dimensional photon echo spectrum of (a) a linear J-aggregate with 36 molecules and of (b-d) various two-dimensional J-aggregates with a brick layer structure of 6x6 molecules. The slip = (b) 0.5, (c) 0.75 and (d) 1.0. Homogeneous broadening with a single Lorentzian line width is assumed and all contours are scaled to the amplitude in the spectrum of the linear aggregate. Negative (bleaching and stimulated emission) peaks are plotted in blue, while positive (induced absorption) peaks are plotted in red and yellow. Contours are at the same absolute level in all panels, and were drawn at -50, -40, -30, -20, 20, 30, 40 and 50 percent of the maximum amplitude of the spectrum of the linear aggregate.}
\end{figure}

We calculated two-dimensional optical spectra using the sum over states method,\cite{Abramavicius.2004.jpcb.108.18034} assuming only homogeneous broadening. This simple method will give a good idea of the peak positions and relative intensities, but not of the details of the line shape. Note that in the calculation of the spectra, we have chosen to vary $J$ in order to obtain similar peak positions for all values of $s$ considered. The reason for this choice is that $J$ is not known a priori in experiment, but the spectral shift with respect to the monomer can be measured. In our spectra, in the case $s=0.5$, because of the presence of both positive and negative couplings in the system, the bright state is not at the bottom of the band. Because of the Pauli exclusion principle, two excitons cannot populate the same state.\cite{Spano.1991.prl.66.1197, Bakalis.1999.jpcb.103.6620} Induced absorption peaks will show up at lower $\omega_3$ than the bleaching and stimulated emission peak. This is most easily understood in the simplified case of a linear aggregate with nearest neighbor interactions only, for which the Hamiltonian can be diagonalized analytically using a Jordan-Wigner transformation.\cite{Kobayashi.1996.book} The two-exciton states are then given as anti-symmetric products of one-exciton states. Because the "first" exciton is not at the bottom of the band, the "second" exciton can go to a lower energy than the first one. This leads to the induced absorption peak. Although this picture of two independent excitons is not strictly valid when long-range interactions are taken into account, the result that an extra induced absorption peak appears at low $\omega_3$ is also found numerically in the full calculation (see Fig.~\ref{fig:spec2D0}).

Note that for a linear aggregate the induced absorption peak is weaker than the bleaching and stimulated emission peak, while the two have approximately equal amplitude for the 2D lattice with $s=1.0$. Both of these systems have the state with largest oscillator strength at the bottom of the band. 
The presence of a large coupling in the y-direction makes the 2D lattice different from a product of one-dimensional aggregates, as considered in Arias et al.\cite{Arias.2012.jpcb.117.4553} We will see that the difference is crucial for a correct determination of the couplings in the system. In this case the bright state is at the bottom of the band, and the two-dimensional spectrum shows a dominant pair of a positive and a negative peak (see Fig.~\ref{fig:spec2D0}). Because the occurrence of positive and negative peaks and their relative intensities in this spectrum are close to the observed spectrum for BIC aggregates,\cite{Arias.2012.jpcb.117.4553} this finding lends support to the model with a slip around $s=0.75-1.0$ for this system. Finally, we note that in the 2D lattice spectrum, a cross peak due to finite size effects is just visible to the right of the main peak.

\subsection{Estimating the nearest neighbour couplings} \label{sec:nncoupl}

We now consider the difference between a linear (N$^2$x1) and a brick layer (NxN) aggregate. In both cases, a dominant pair of positive and negative peaks is found in the 2D spectrum.\cite{Stiopkin.2006.jpcb.110.20032, Arias.2012.jpcb.117.4553} 

The relation between the nearest neighbor coupling and the peak shift of the aggregate compared to the monomer, which can be measured by finding the maximum in the linear spectra, is different. This simple method, which relies on the fact that excitonic coupling shifts the peak, is frequently used to determine the coupling from experimentally obtained spectra, even though it neglects the shift in the single molecule transition frequency due to the electrostatic effect of the different environment in both cases. The difference is easily explained from the fact that there are more neighbors and that there is therefore more coupling in the 2D aggregate. Quantitatively, the peak shift in a linear aggregate is 2.4 times the nearest neighbor coupling.\cite{Fidder.1991.jcp.95.7880} In the 2D system with $s=1.0$, we find that the shift is several times larger ($4.9 J$ for a 6x6 bricklayer lattice). Thus, while estimating the couplings from the spectral shift, it is important to take the aggregate geometry into account. Here, we established the rule that can be used to estimate the coupling in a two-dimensional brick layer aggregate from the measured linear absorption spectrum for parameters $s=1$, $A=2$. For other parameters a similar rule can be established, which will, in general, be different from the rule derived from calculations on linear aggregates. We note that the magnitude of the couplings and, therefore, spectral shifts strongly depend on the aggregate under study. In particular, in Ref.~\onlinecite{Muller.2013.jcp.139.044302} much smaller couplings were found than in Ref.~\onlinecite{Arias.2012.jpcb.117.4553}. However, the relation between spectral shift and coupling does not depend on the absolute value of these numbers. We note that the practical rule described here is related to well established results derived from sum rules.\cite{Merrifield.1968.jcp.48.3693}

\subsection{Population relaxation} \label{sec:lifetime}

We next consider the life time of the bright state, which for both the linear aggregate and the brick layer lattice with $A=2$ and $s=1.0$ lies at the bottom of the band. This life time can be measured experimentally using pump-probe spectroscopy. For the well studied system of linear aggregates in a glass, the temperature dependence of the life time of this state is mostly determined by scattering to higher states with the absorption of phonons. The dephasing associated with population relaxation is found to dominate, leading to a $T^{3.5}$ dependence of the homogeneous line width.\cite{Heijs.2005.jcp.123.144507} This result is obtained from perturbation theory, where the scattering rate between two eigenstates is the product of three contributions: the Boltzmann factor of the phonons $n$ evaluated at the energy gap between both states, the exciton-phonon spectral density evaluated at the energy gap, which is taken cubic for the glass environment, and the overlap of the wave functions of the two states. We refer to Ref.~\onlinecite{Heijs.2005.jcp.123.144507} for further details. In this model, it is clear that the density of states at the band edge is very important for the life time.

As was shown before, this density of states is very different in linear and 2D aggregates. As a result, the established theory for linear aggregates in a glass must be used with caution here. The Boltzmann factor is much larger and more strongly peaked in the linear chain, leading to larger scattering rates, and a smaller life time. Although the spectral density typically increases with energy, this is counteracted by the stronger exponential decay of the Boltzmann factor. We also expect different behavior of the life time as a function of temperature, because the argument used in Heijs et al. to arrive at the $T^{3.5}$ dependence is valid only for $kT > E_2 - E_1$. We therefore expect a different behaviour of the temperature dependence of the homogeneous line width, irrespective of the details of the interaction with the phonons.

\begin{figure*}[t]
 \includegraphics{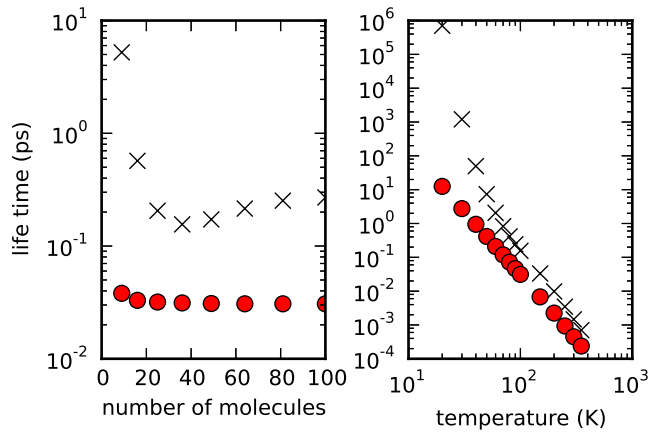}
\caption{\label{fig:lifetime} Calculated intraband scattering life time of the bright exciton state with the model of Heijs et al. in an NxN brick layer lattice (crosses, $A=2, s=1$) and an N$^2$x1 linear chain (circles) (left) as a function of the system size ($N^2$) at a temperature of 100 K and (right) as a function of temperature for $N=6$. The coupling in the x-direction is $J=-500 \mathrm{cm}^{-1}$.}
\end{figure*}

As the numerical calculations presented in Fig.~\ref{fig:lifetime} show, the life time is up to more than an order of magnitude larger in the brick layer system than in the linear aggregate for the same system (delocalization) size. Life times were computed for the same cubic spectral density in all cases and, in contrast to the previous sections, for a nearest neighbour coupling strength of $J=-500 \mathrm{cm}^{-1}$, which is a typical value for the aggregates considered. All long range couplings were included as well. We note that the population relaxation rate as a function of temperature exhibits a power law dependence for linear aggregates, while a kink is observed for a 2D lattice. The kink can be understood as follows. In the argument leading to a power law dependence of the scattering rates as a function of temperature, the summation over discrete states is replaced by an integration.\cite{Heijs.2005.jcp.123.144507} This replacement is valid if the temperature is large compared to the energy gap between the relevant exciton states, which are the lowest two states in the band. For the 10x10 brick layer lattice used in Fig.~\ref{fig:lifetime}, the energy gap between these states is found to be 71 cm$^{-1}$, which corresponds to a temperature of roughly 100 K. Therefore, the power law dependence, which holds for high temperatures compared to the energy gap, breaks down and a kink is observed in the life time. We also note that the temperature dependence of the life time depends very strongly on $A$ and $s$. As discussed in the appendix, there is almost no temperature  dependence for certain values, while for other values the life time varies over orders of magnitude.

Note that, to make a direct comparison possible, these calculations assume that the spectral density is the same for linear and two-dimensional aggregates. We will argue later, based on the analysis of the experimentally measured temperature dependence of the homogeneous line width, that there is a linear component in the spectral density for the two-dimensional aggregates. However, the suppression of relaxation rates in the two-dimensional aggregates based on the much smaller Boltzmann factor will occur irrespective of the spectral density.

\begin{figure*}[t]
 \includegraphics{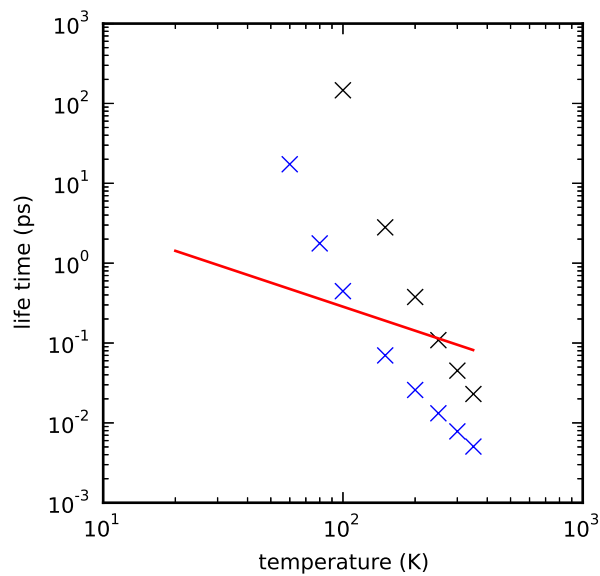}
\caption{\label{fig:lifetimelinJ} Population life time (crosses) and pure dephasing life time (line) as a function of temperature for a linear spectral density in a 2D aggregate. Parameters for the solid line, which shows the pure dephasing life time, have been extracted from the BIC experiment (see Sec. \ref{sec:exp} for details), from which we can estimate the slope of the pure dephasing contribution as a function of temperature, but not the delocalization size. For the calculation of the dephasing due to population relaxation, which is plotted as crosses, the size is used as an input parameter. Calculations were performed for two different sizes, and the results are plotted with black crosses are for a delocalization size of $N_k=9$ molecules (3x3 brick layer lattice), while blue crosses are for $N_k=25$ molecules (5x5 brick layer lattice).}
\end{figure*}

If, indeed, a linear spectral density is more appropriate for two-dimensional aggregates,
we can also obtain population relaxation rates for this case. The result is shown in figure \ref{fig:lifetimelinJ}. Parameters were estimated based on the experimental data for a BIC aggregate, see section \ref{sec:exp}. For this choice of the spectral density, the pure dephasing can be characterized by a rate if the temperature is low compared to the cut-off frequency of the bath.\cite{Reichman.1996.jcp.105.10500} The life time is also plotted in the figure. We observe that pure dephasing dominates the homogeneous line width up to a certain temperature, which depends on the system size used in the simulations (i.e. the exciton localization size). For higher temperatures, population relaxation becomes more important. This finding can be used to determine an upper limit for the delocalization size.

\subsection{Dephasing in two-dimensional aggregates} \label{sec:linewidth}

Comparison with experiment shows that the predominance of dephasing from population relaxation, which is found for linear aggregates, does not explain experiment for the thin film. A perfectly linear relation between the homogeneous line width and the temperature was measured.\cite{Arias.2012.jpcb.117.4553} Although this finding could be explained from population relaxation with a different $\omega$ dependence of the spectral density, this is not a plausible explanation of the experimental findings. A sub-Ohmic power\cite{Arias.2012.jpcb.117.4553}, on the order of $\omega^{0.5}$ in the spectral density would be needed. There is no clear microscopic mechanism that would lead to this behavior. Furthermore, it would be very surprising that the temperature dependence of the homogeneous line width is exactly linear. Therefore, it is more logical to assume that the mechanism leading to homogeneous broadening is different in linear and 2D aggregates. 

The different mechanism can be understood by the difference in population relaxation rates found in the previous section. In linear aggregates, the density of states at the band edge is large, leading to strong scattering between exciton states and fast relaxation of the population in the bright state. This population relaxation leads to dephasing, which dominates the homogeneous line width. In contrast, in two-dimensional aggregates, the density of states at the band edge is orders of magnitude smaller. Population relaxation, which depends strongly on the difference in energy between eigenstates, is suppressed for the smooth spectral densities that we assume here. The accompanying dephasing is therefore also much weaker than in the case of linear aggregates. Therefore, pure dephasing becomes more important and can dominate the homogeneous line width. This is the main finding of this paper. This model predicts that the life time broadening is only a small part of the homogeneous line width. This prediction could be tested experimentally by measuring the life time of the bright state with pump-probe spectroscopy. 

We will now see how the dominance of pure dephasing in two-dimensional aggregates leads to a linear scaling of the homogeneous line width with temperature.
We assume that the spectral density is polynomial in frequency up to a cut-off frequency $\omega_C$,
\begin{equation}
  J(\omega) = C_J \omega^\alpha f_J(\omega/\omega_C),
\end{equation}
where $f_J(x)$ is a cut-off function. Common forms for this function are an exponential, a Lorentzian or a Heaviside step function. While this spectral density quite generally describes the interaction with the bath, we assume here that intramolecular vibrations do not play a role.

We are now in a position to analyse the pure dephasing term $g(t)$. 
In the fast modulation limit, which we assume here because, as we will see, it leads to a linear relation between the homogeneous line width and temperature, one can replace the correlation function by a delta function, $L(t) = \Gamma^\mathrm{pure} \delta(t)$. The pure dephasing contribution to the spectrum is then given by a rate $\Gamma^\mathrm{pure}$, because $g(t) = \Gamma^\mathrm{pure} t$. By comparing with Eq.~\ref{cfj} one finds that the rate is related to the slope of the spectral density at zero frequency,\cite{Weiss.2008.book, Kreisbeck.2012.jpcl.3.2828} $\Gamma^\mathrm{pure} = \lim_{\omega \to 0} J(\omega)/\beta \omega$. Therefore, there is no pure dephasing rate for a super-Ohmic spectral density ($\alpha>1$). Note that this statement is also valid outside the fast modulation limit, because the linear term in the line shape function is also given by the slope of the spectral density at zero frequency in the more general case. There is still pure dephasing, but $g(t)$ has no linear term. For an Ohmic spectral density ($\alpha=1$), which is the most commonly used form, because it corresponds to a linear density of states and a frequency independent exciton phonon coupling, we see that the pure dephasing rate is linearly proportional to temperature.\cite{Mukamel.1995.book} Thus, for an Ohmic spectral density in the fast modulation limit, we expect a linear scaling of the homogeneous line width with temperature. We note that it would be desirable to measure the time scale of the bath directly to strengthen this argument.

To close this section, we briefly discuss Kubo stochastic line shape theory,\cite{Kubo.1954.jpsj.9.935} which is valid in the high temperature limit, to estimate the temperature dependence of the pure dephasing in the slow dephasing limit. The correlation function in this case is given by
\begin{equation}
L(t) = \sigma^2 e^{-\omega_C t},
\end{equation}
where the variance of the fluctuations can be expressed in terms of the reorganization energy $\lambda$ and temperature by calculating the correlation function from the Drude-Lorentz spectral density $J(\omega) = 2 \lambda \omega_C \omega / (\omega^2 + \omega_C^2)$. One finds $\sigma^2 = 2 \lambda/\beta$. The lineshape function for this model is easily calculated to be
\begin{equation} \label{DLgt}
g(t) = \frac{\sigma^2}{\omega_C^2}(\omega_C t - 1 + e^{-\omega_C t}).
\end{equation}
One sees that in the fast modulation limit $g(t) = 2 \lambda t / \beta \omega_C$, in agreement with the pure dephasing rate introduced earlier. In the slow modulation limit, $g(t) = \sigma^2 t^2/2$, where $\sigma$ is related to the inhomogeneity in the energies of exciton states. The line shape in the frequency domain is a Gaussian with standard deviation proportional to $\sigma$. Because the variance $\sigma^2$ is linearly proportional to temperature, the line width scales as the square root of temperature in this regime. Note that this regime is less relevant for our discussion of the homogeneous line width, which is determined by fast fluctuations, while very slow fluctuation contribute only to the inhomogeneous line width. However, this analysis shows that if the fast modulation limit is not strictly applicable,\cite{Jang.2002.jpcb.106.8313} deviations from linear scaling of the line width with temperature are expected.

Before presenting a comparison to experiment, we briefly discuss the role of static disorder.  As indicated by Eq.~\ref{DLgt}, at low temperature, the line-shape and its T-dependence will be dominated by inhomogeneous broadening resulting from static disorder. In fact, a published calculation [see Fig. 7 of Ref.~\onlinecite{Moix.2013.njp.15.085010}] clearly shows the transition from inhomogeneous broadening to homogenous broadening in the disordered chain system. Interestingly, this transition corresponds to optimal diffusion along the chain, suggestion optimization when dynamics and static disorders balance.  Another point is that the relative contribution of pure dephasing decreases with the size of localization, as shown in Fig.~\ref{fig:lifetimelinJ}.  It is known that the Anderson localization size scales with the disorder and this universal scaling depends critically on the dimensionality.  This scaling and its implication on diffusion were reported in a recent study \cite{Chuang.2014.jpcb.118.7827} and will be further explored in two-dimensional J-aggregates.\cite{Cherninprep}

\section{Extracting parameters from experiment} \label{sec:exp}

By comparing the measured 2D spectra in ref. \onlinecite{Arias.2012.jpcb.117.4553} with our calculated spectra, we observe that a slip around 1.0 is the most plausible structure for these systems. A slip around 0.5 and smaller is ruled out, because an extra induced absoption peak appears below the bleaching and stimulated emission peak, which is not observed in experiment.

As explained in section \ref{sec:nncoupl}, the estimate of the resonant transfer interactions in two-dimensional aggregates should be done with a different formula than in linear aggregates. To obtain an estimate of the coupling $J$, one should divide the excitonic peak shift by a factor of 4.9. This factor depends on $s$ and $A$. By doing this, we find that the nearest neighbor coupling in the aggregate studied by Arias et al. is around -75 meV for BIC (instead of -153 meV based on a linear model) and -104 meV for U3 (instead of -212 meV). In PTCDA aggregates considered by Mueller et al., the shift was found to be 400 cm-1. This would lead to an interaction of -82 cm$^{-1}$. This value corresponds to -10 meV, and is therefore significantly smaller than for the other two aggregates. 

Under the assumption that the homogeneous line width is completely determined by pure dephasing, we can extract system parameters from the experimental data by following the discussion in the previous section.  
The slope of the homogeneous line width versus temperature, $\Gamma = S T$, is given by $S=2 \lambda k_B/ \omega_C N_k$, where $N_k$ is the localization size of the exciton. Thus, we find that $2 \lambda / \omega_C = 0.268 N_k$ for BIC. To proceed further, it would be desirable to measure $\omega_C$ independently, for example from the time dependent Stokes shift. 

From this estimate, by assuming a value for $N_k$, we can derive the parameters for the model by Heijs et al., and calculate the population relaxation times as shown in figure \ref{fig:lifetimelinJ}. We used the same large value of $\omega_C = 10 J$, which means that we are consistently in the Markovian regime. We conclude that in order to obtain a linear relationship between homogeneous life time and temperature up to 250 K, the exciton should be localized on a segment smaller than 9 molecules (3x3). For larger temperatures, deviations from linear behaviour are expected, as can be seen in the figure. This is consistent with the dynamic localization size estimated from experiments on BIC and U3, \cite{Arias.2012.jpcb.117.4553} and slightly smaller than values found for PTCDA.\cite{Muller.2013.jcp.139.044302} Note that in our modified Redfield approach dynamic localization is not included by construction and that we therefore regard $N_k$ as a parameter in the calculations.

\section{Conclusion} \label{sec:conc}

In conclusion, we have used the standard Frenkel exciton model to study the excitonic properties in brick layer thin film J-aggregates. We have introduced a novel theory to explain the experimentally measured linear temperature dependence of the pure dephasing rate.

We have found that the exciton couplings determined from peak shifts in the linear spectrum depend strongly on the geometry of the system. In particular, the couplings in two-dimensional aggregates are a few times smaller than estimates based on a linear aggregate model.

To explain the linear scaling of the homogeneous line width with temperature, we propose pure dephasing as the main homogeneous broadening mechanism. Because the energy gaps between exciton states at the bottom of the band are much larger in two-dimensional than in linear aggregates, population relaxation is surpressed. This leads to a smaller contribution to the line width from population relaxation in two-dimensional aggregates, as well as to longer exciton life times. From the experimental data, we can extract the product of the reorganization energy and the typical bath time reorganization time scale. Pump-probe experiments, which can measure the lifetime of excited states, are suggested to confirm whether dephasing due to population relaxation is indeed less important than pure dephasing. It will also be valuable to determine the time scale of the reorganization of the phonon environment experimentally. 

A much simpler model could in principle explain the exact linear temperature dependence. Stochastic line shape theory for a single two-level system coupled to a bath in the fast modulation limit predicts a Lorentzian homogeneous line with a width $\Delta^2 \tau$, where $\Delta$ is the standard deviation of the fluctuations and $\tau$ their correlation time. Because the variance of the fluctuations depends linearly on temperature this would explain the observed temperature dependence irrespective of the form of the spectral density. 

A microscopic model that could make the exciton state behave as an effective two-level system is the presence of correlated fluctuations. If the site energy fluctuations are not independent, as in the model of Heijs et al., but correlated over a distance comparable to the exciton localization size, they will not lead to scattering between eigenstates, and therefore no contribution to the line width from scattering in the exciton band occurs. We believe that the localization size is large enough to make the model of correlated fluctuations not plausible.

Our calculations are limited by the assumption in modified Redfield theory, where off-diagonal fluctuations in the exciton basis are treated perturbatively and therefore dynamic localization is not included. They could be improved by taking both the diagonal and off-diagonal system bath coupling terms into account non-perturbatively. This could be achieved with the cumulant expansion method,\cite{Ma.2015.jcp.142.094106, Ma.2015.jcp.142.094107} stochastic path integrals,\cite{Moix.2015.jcp.142.094108} or hierarchy of equation of motion simulations.\cite{Tanimura.2006.jpsj.75.082001} These methods can be used to study the effect of dynamic localization,\cite{Moix.2012.prb.85.115412} but are difficult to apply given the large reorganization energy and low temperature compared to the excitonic coupling in the system. They could also be used to calculate the localization size, which is treated as a parameter in our work. We remark that modified Redfield theory becomes more accurate with an increase in the excitonic energy gap, which means that the theory is more accurate for systems of modest localization size, as we consider here. For some aggregates, intermolecular vibrations are important,\cite{Muller.2013.jcp.139.044302, Stradomska.2010.jcp.133.094701} which could be considered in future work. Experimental measurement of the Stokes shift may help to improve our understanding of the role of vibrations. Finally, it would be desirable to get a better estimate of the resonant transfer interactions from quantum chemical simulations and to include radiative decay to the ground state in our calculations. On the experimental side, collecting 2D spectra as a function of the waiting time to elucidate energy transfer mechanisms\cite{Schlau.2009.jpcb.113.15352} would be a possible future extension.

The results presented here contribute to the understanding of the photophysics of two-dimensional J-aggregate thin films. 

\section*{Acknowledgement}
Work by KAN and AGD was supported as part of the Center for Excitonics, an Energy Frontier Research Center funded by the U.S. Department of Energy, Office of Science, Basic Energy Sciences under Award \# DE-SC0001088. JC was supported by the NSF (grant no. CHE-1112825). AGD was furthermore supported by a Marie Curie International Incoming Fellowship within the 7th European Community Framework Programme (grant no. 627864). HGD acknowledges financial support by the Joachim-Herz-Stiftung, Hamburg within the PIER Fellowship program.

\section*{Appendix: more geometries}
In this appendix, we consider a wider range of geometries than were presented in the main text, in particular, a brick layer lattice with a larger aspect ratio $A=3$. 

\begin{figure*}[t]
 \includegraphics{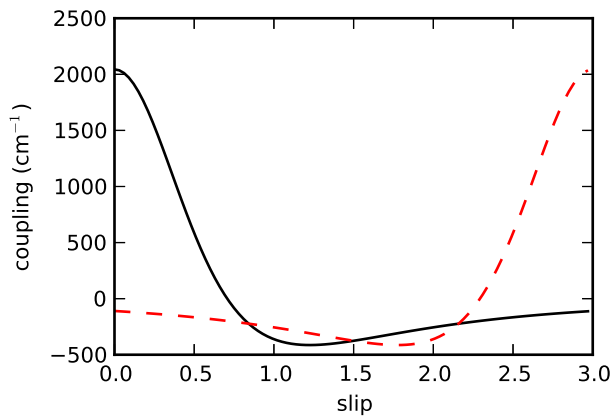}
\caption{\label{fig:couplingA3} Resonant transfer interaction as a function of the slip $s$ for $A=3$. All other parameters are the same as in Fig.~\ref{fig:coupling}.}
\end{figure*}

\begin{figure*}[t]
 \includegraphics{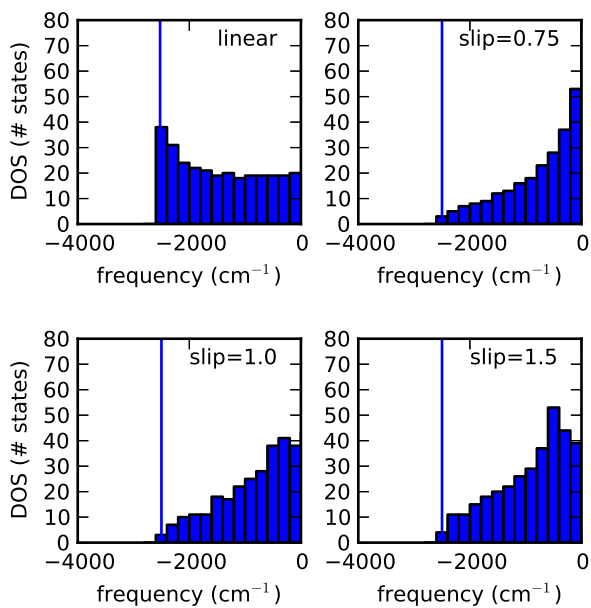}
\caption{\label{fig:dosA3} Density of states for $A=3$ and various values of $s$. The bin size is 200 cm$^{-1}$. }
\end{figure*}

In figure \ref{fig:couplingA3} we plot the resonant transfer interaction for an aspect ratio of 3. Similar as in the case $A=2$, we find that there is only a limited range of values of the slip $s$ for which interactions are negative in the horizontal as well as the vertical direction. In figure \ref{fig:dosA3} we plot the density of states for $A=3$. We observe a clear difference in the density of states at the band edge between linear aggregates and brick layer aggregates. We note that for a slip of $s=0.5$, the state with maximum oscillator strength is blue shifted with respect to the monomer. 

\begin{figure*}[t]
 \includegraphics{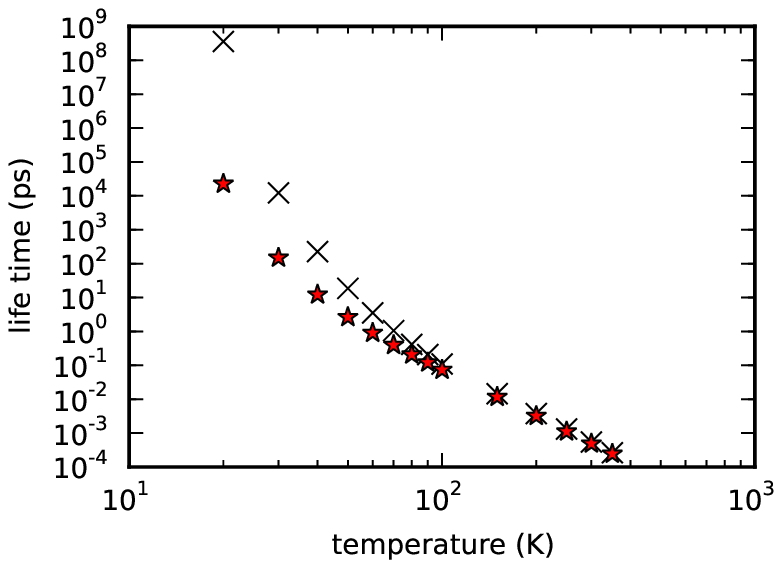}
\caption{\label{fig:lifetimeA3} Calculated intraband scattering life time with the model of Heijs et al. in an NxN brick layer lattice (crosses: $A=3, s=1$, stars: $A=3$, $s=1.5$) as a function of temperature for $N=6$. The coupling in the x-direction is $J=-500\,\mathrm{cm}^{-1}$.}
\end{figure*}

We calculated the life time of the bright state for a $6x6$ brick layer lattice for different values of $A$ and $s$, using the same procedure as outlined in the main text. For $A=2$, with $s=0.25$ and $s=0.5$, the temperature dependence varies dramatically less than for the value of $s=1.0$ used in the main text (data not shown). In particular, for $s=0.25$ the lifetime is constant at the very small value of 1.5 $\cdot\, 10^{-7}$ ps over the entire range of temperatures considered (20 to 350 K). For $s=0.5$, the life time decreases from 8 $\cdot\, 10^{-4}$ ps at 20 K to 2 $\cdot\, 10^{-4}$ ps at 350 K, still a much smaller variation than was observed in Fig.~\ref{fig:lifetime} for $s=1.0$. We attribute the difference to the presence of states which are lower in energy than the bright state. The system can relax to these states by spontaneous emission of a phonon, which is independent of temperature.

For $A=3$ and $s=0.5$, we also find a constant lifetime over the range of temperature from 20 K to 350 K. For $s=1.0$ and $s=1.5$, we find a very strong dependence on temperature, as can be seen from the data plotted in Fig.~\ref{fig:lifetimeA3}. We conclude that measuring the temperature dependence of the life time of the bright state can be used as a tool to probe the molecular arrangement in brick layer aggregates.

\end{document}